\documentclass[conference,compsoc]{IEEEtran}
\IEEEoverridecommandlockouts

\usepackage{amsmath,amsfonts,amssymb,amsthm,bm}
\usepackage{xcolor}
\usepackage{setspace}
\usepackage[noend]{algpseudocode}
\usepackage[normalem]{ulem}

\usepackage{cite}
\usepackage{graphicx}
\usepackage{textcomp}
\usepackage[caption=false,font=footnotesize,labelfont=sf,textfont=sf]{subfig}
\def\BibTeX{{\rm B\kern-.05em{\sc i\kern-.025em b}\kern-.08em
    T\kern-.1667em\lower.7ex\hbox{E}\kern-.125emX}}

\usepackage[ruled,norelsize]{algorithm}
\makeatletter
\newcommand{\removelatexerror}{\let\@latex@error\@gobble}
\makeatother 

\begin{document}

\title{A GPU-Accelerated Barycentric Lagrange Treecode\\
\thanks{Supported by National Science Foundation grant DMS-1819094, 
Extreme Science and Engineering Discovery Environment (XSEDE) grants ACI-1548562 and ASC-190062, and the Mcubed program and Michigan Institute for Computational Discovery and Engineering (MICDE) at the University of Michigan. Nathan Vaughn and Leighton Wilson are co-first authors.
\copyright~2020 IEEE.  Personal use of this material is permitted.  Permission from IEEE must be obtained for all other uses, in any current or future media, including reprinting/republishing this material for advertising or promotional purposes, creating new collective works, for resale or redistribution to servers or lists, or reuse of any copyrighted component of this work in other works.}}

\author{\IEEEauthorblockN{Nathan Vaughn}
\IEEEauthorblockA{\textit{Department of Mathematics} \\
\textit{University of Michigan}\\
Ann Arbor, USA \\
njvaughn@umich.edu}
\and
\IEEEauthorblockN{Leighton Wilson}
\IEEEauthorblockA{\textit{Department of Mathematics} \\
\textit{University of Michigan}\\
Ann Arbor, USA \\
lwwilson@umich.edu}
\and
\IEEEauthorblockN{Robert Krasny}
\IEEEauthorblockA{\textit{Department of Mathematics} \\
\textit{University of Michigan}\\
Ann Arbor, USA \\
krasny@umich.edu}
}

\maketitle

\begin{abstract}
We present an MPI + OpenACC implementation of the kernel-independent 
barycentric Lagrange treecode (BLTC) for
fast summation of particle interactions on GPUs.
The distributed memory parallelization 
uses recursive coordinate bisection for domain decomposition
and
MPI remote memory access to build locally essential trees on each rank.
The particle interactions are organized into
target batch/source cluster interactions which efficiently map onto the GPU;
target batching provides an outer level of parallelism,
while the direct sum form of the barycentric particle-cluster
approximation
provides an inner level of parallelism.
The GPU-accelerated BLTC performance is demonstrated on several test cases 
up to 1~billion particles interacting via the Coulomb potential and Yukawa potential.
\end{abstract}

\begin{IEEEkeywords}
Heterogeneous (hybrid) systems, 
Graphics processors,
Load balancing and task assignment,
Interpolation,
Numerical algorithms,
Parallel algorithms,
Chebyshev approximation and theory,
Integral Equations
\end{IEEEkeywords}

\section{Introduction}
Calculation of long-range particle interactions is essential in 
many areas of computational physics,
for example in computing electrostatic or gravitational potentials and forces. 
In a system with $N$ particles,
the cost of direct summation scales like $O(N^2)$ 
which is prohibitively slow for large systems,
but improved hardware can reduce the cost.
For example,
direct summation has been implemented on graphics processing units (GPUs) 
with 25$\times$ speedup over an optimized CPU implementation~\cite{Elsen:2006aa},
and 
250$\times$ speedup over a portable C implementation~\cite{Nyland:2009aa}.
However the GPU implementation of direct summation 
does not improve the scaling with system size $N$.
To address this issue several hierarchical fast summation methods 
with subquadratic scaling are available 
including the Barnes--Hut treecode \cite{Barnes:1986aa}
and
Greengard--Rokhlin fast multipole method (FMM) \cite{Greengard:1987aa};
the price however is that these methods are more complex than
direct summation,
and
their parallel implementation is a topic of ongoing research
especially as HPC architectures evolve.   

{\bf GPU-accelerated hierarchical methods.}
There is a large and growing body of work on GPU-accelerated 
hierarchical fast summation methods
and
here we describe only a brief sample.
Hamada et al. introduced CUDA implementations of a treecode for gravitational $N$-body simulations 
and 
an FMM for turbulence simulations that ran on 256 GPUs~\cite{Hamada:2009aa}. 
B\'edorf et al. developed a gravitational $N$-body treecode called Bonsai
running entirely on GPUs,
which enabled a galaxy simulation on 
18600 GPUs~\cite{Bedorf:2012aa,Bedorf:2014aa}.
Burtscher and Pingali presented a CUDA treecode
which replaced the pointer-chasing recursion used in many CPU treecodes 
with an iteration over array structures~\cite{Burtscher:2011aa}. 
Yokota and Barba implemented a GPU treecode and FMM
with multipole expansions to simulate vortex ring dynamics~\cite{Yokota:2011aa}.
Fortin and Touche used a 
dual tree traversal scheme on GPUs for astrophysical $N$-body 
simulations~\cite{Fortin:2019aa}.

{\bf Kernel-independent methods.}
Many hierarchical fast summation methods rely on 
analytic approximations specific to a given kernel;
examples include multipole expansions
and
exponential representations
in FMMs for the Coulomb and Yukawa potentials~\cite{Cheng:1999aa,Greengard:2002aa},
and
Cartesian Taylor expansions 
in treecodes~\cite{Duan:2001aa,Li:2009aa}.
There is also interest in kernel-independent methods suitable
for a large class of kernels;
such methods require only kernel evaluations,
rather than analytic approximations specific to each kernel.
Among these,
the kernel-independent FMM (KIFMM) uses equivalent densities to approximate
well-separated particle interactions
and
has been parallelized for heterogeneous architectures 
using OpenMP and CUDA 
\cite{Ying:2003aa,Ying:2004aa,Lashuk:2012aa, March:2015aa, Malhotra:2015aa, Malhotra:2016aa},
while
the black-box FMM (bbFMM) 
uses polynomial interpolation and SVD compression
and
has been ported to GPUs using CUDA \cite{Fong:2009aa,Takahashi:2012aa,Agullo:2016aa}.


{\bf Present work.}
A key goal in designing hierarchical fast summation methods is to achieve
good parallel scaling as well as sufficient accuracy required by applications.
With this in mind,
we present an MPI + OpenACC implementation of the recently developed
barycentric Lagrange treecode (BLTC)~\cite{Wang:2019aa}. 
The scheme approximates well-separated particle-cluster interactions
using barycentric Lagrange interpolation at Chebyshev points of the second kind,
which is stable and efficient~\cite{Berrut:2004aa}.
The BLTC is kernel-independent
and
the structure enables an efficient MPI + OpenACC implementation
on multiple GPU nodes,
using recursive coordinate bisection for domain decomposition
and
MPI remote memory access to build locally essential trees on each rank.
The particle interactions are organized into target batch/source cluster interactions
which efficiently map onto the GPU;
target batching provides an outer level of parallelism,
while the direct sum form of the barycentric particle-cluster approximation
provides an inner level of parallelism.

Numerical results are presented for the Coulomb potential and Yukawa potential.
First we consider a system with 1~million particles
and
compare the BLTC running on a single GPU versus a 6-core CPU,
with treecode parameters set to span the range 
between 2 digit accuracy and machine precision.
Then with parameters set to achieve 5-6 digit accuracy,
we demonstrate
weak scaling up to 1.024~billion particles
and
strong scaling for 64~million particles, from 1 to 32 GPUs (8 nodes).

The remainder of the paper is organized as follows.
Section~\ref{section:bary-treecodes} 
describes the barycentric Lagrange treecode.
Section~\ref{section:implementation-details} 
describes the MPI + OpenACC distributed memory implementation.
Section~\ref{section:results} 
presents numerical results.
Section~\ref{section:conclusion} gives the conclusions.


\section{Barycentric Lagrange Treecode}
\label{section:bary-treecodes}

Consider the problem of evaluating the 
electrostatic potential due to a set of charged particles, 
\begin{equation}
    \varphi(\mathbf{x}_i) = \sum_{j=1}^N G(\mathbf{x}_i,\mathbf{y}_j)q_j, 
    \quad i=1:N,
    \label{eqn:particle-particle}
\end{equation}
where $G(\mathbf{x}_i,\mathbf{y}_j)q_j$ is the interaction between
a target particle ${\bf x}_i$
and
a source particle ${\bf y}_j$ with charge $q_j$.
In this work the kernel $G(\mathbf{x},\mathbf{y})$ is either the
Coulomb potential or the Yukawa potential,
\begin{equation}
    G(\mathbf{x},\mathbf{y}) = \frac{1}{|{\bf x}-{\bf y}|}, 
    \quad
    G(\mathbf{x},\mathbf{y}) = 
    \frac{e^{-\kappa|{\bf x}-{\bf y}|}}{|{\bf x}-{\bf y}|},
    \label{eqn:kernels}
\end{equation}
where $\kappa$ is the inverse Debye length,
but in general it can be any non-oscillatory kernel that is smooth 
for ${\bf x} \ne {\bf y}$.
Expressions similar to~\eqref{eqn:particle-particle}
also arise in gravitational simulations where the particles are point masses
and 
in boundary element methods 
where the particles are quadrature points of a discretized convolution integral.
Computing the potentials by direct summation requires $O(N^2)$ operations,
and
the barycentric Lagrange treecode described 
below computes approximations in $O(N\log N)$ operations 
in a way that enables an efficient GPU implementation.


\subsection{Barycentric Lagrange Interpolation}
\label{section:lagrange-form}

The BLTC is based on the barycentric Lagrange form of the 
interpolating polynomial,
which we briefly review in 1D.
Given a function $f(x)$ evaluated at $n+1$ points
$s_k$, 
the Lagrange form of the interpolating polynomial is
\begin{equation}
    p_n(x) = \sum_{k=0}^n f(s_k) L_k(x),
\end{equation}
where $L_k(s_j) = \delta_{jk}$.
The Lagrange polynomials can be expressed in various bases,
but Berrut and Trefethen~\cite{Berrut:2004aa}
advocated in favor of the barycentric form,
\begin{equation}
L_k(x) = \frac{\displaystyle 
\frac{w_k}{x-s_k}}
{\displaystyle \sum_{k^\prime=0}^n \frac{w_{k^\prime}}{x-s_{k^\prime}}}, 
\quad w_k = \frac{1}{\prod_{j=0,j\ne k}^n(s_k-s_j)},
\label{eqn:1d-barycentric}
\end{equation}
for $k=0:n$.
Note that this expression for $L_k(x)$
has removable singularities at the interpolation points,
\begin{equation}
\lim_{x \to s_{k^\prime}}L_k(x) = 
\frac{w_k}{w_{k^\prime}}\lim_{x \to s_{k^\prime}}
\frac{x-s_{k^\prime}}{x-s_k} = \delta_{kk^\prime}.
\end{equation}
In practice the removable singularities do not pose a problem
and
Section~\ref{section:interpolation-singularities} will discuss how they
are handled in the code~\cite{Berrut:2004aa,Wang:2019aa}.

The BLTC uses Chebyshev points of the second kind due to their good 
approximation properties~\cite{Salzer, Berrut:2004aa}.
For the interval $[-1,1]$, these points are given by
\begin{equation}
s_k = \cos\theta_k, \quad \theta_k = \pi k/n, \quad k = 0:n,
\end{equation}
and their corresponding interpolation weights are given by
\begin{equation}
    w_k = (-1)^k\delta_k,
    \label{eqn:barycentric_weights}
\end{equation}
where $\delta_k=1/2$ if $k=0$ or $n$, and $\delta_k=1$ otherwise.
For a different interval $[a,b] \ne [-1,1]$,
the Chebyshev points $s_k$ can be linearly mapped
and
the barycentric weights~\eqref{eqn:barycentric_weights} stay the same.

In 3D, we consider a target particle $\mathbf{x} = (x_1,x_2,x_3)$ 
and source particle $\mathbf{y} = (y_1,y_2,y_3)$.
In many cases of interest,
including the Coulomb and Yukawa potentials used in this work,
the kernel $G(\mathbf{x},\mathbf{y})$ 
is smooth for $\mathbf{x} \ne \mathbf{y}$
and 
can be approximated locally with a polynomial.
Here we employ a tensor product of $(n+1)^3$ 
Chebyshev grid points $\mathbf{s_k}=(s_{k_1}, s_{k_2}, s_{k_3})$ 
and
interpolate with respect to the source variable,
\begin{equation}
    G(\mathbf{x},\mathbf{y}) \approx \sum_{\bf k} G(\mathbf{x},\mathbf{s_k}) L_{k_1}(y_1)L_{k_2}(y_2)L_{k_3}(y_3),
    \label{eqn:kernel-barycentric-lagrange}
\end{equation}
where the sum over ${\bf k} = (k_1,k_2,k_3)$ 
is performed in each index $k_\ell=0:n, \ell=1,2,3$ for interpolation degree $n$. 


\subsection{Particle-Cluster Interactions}

Next consider a target particle $\mathbf{x}_i$ interacting with a 
cluster of source particles $C = \{{\bf y}_j\}$,  
where $\mathbf{y}_j$ has coordinates $(y_{j1},y_{j2},y_{j3})$ and charge $q_j$.
The potential at $\mathbf{x}_i$ due to the particle-cluster interaction is given by
\begin{equation}
    \varphi(\mathbf{x}_i,C) = \sum_{\mathbf{y}_j\in C} G(\mathbf{x}_i,\mathbf{y}_j)q_j,
    \label{eqn:particle-cluster-exact}
\end{equation}
and
using the kernel approximation~\eqref{eqn:kernel-barycentric-lagrange}, 
this can be approximated by 
\begin{equation}
    \varphi(\mathbf{x}_i,C) \approx 
    \sum_{\mathbf{y}_j\in C} \sum_{\bf k} G(\mathbf{x}_i,\mathbf{s_k})
    L_{k_1}(y_{j1})L_{k_2}(y_{j2})L_{k_3}(y_{j3})q_j.
    \label{eqn:particle-cluster-approx}
\end{equation}
Changing the order of summation yields 
\begin{equation}
    \varphi(\mathbf{x}_i,C) \approx  
    \sum_{\bf k} G(\mathbf{x}_i,\mathbf{s_k})\widehat{q}_{\bf k},
    \label{eqn:particle-cluster-approx-rearranged}
\end{equation}
where $\widehat{q}_{\bf k}$, the modified charges, are given by
\begin{equation}
    \label{eqn:modified-weights}
    \widehat{q}_{\bf k} = \sum_{\mathbf{y}_j\in C} L_{k_1}(y_{j1})L_{k_2}(y_{j2})L_{k_3}(y_{j3})q_j. 
\end{equation}
There are two important consequences of this rearrangement.
First, each $\widehat{q}_{\bf k}$ is independent of the target particle ${\bf x}_i$ 
and 
can be precomputed, stored, and reused for all targets 
interacting with this cluster.
Second, 
the particle-cluster approximation~\eqref{eqn:particle-cluster-approx-rearranged} 
has the same direct sum form
as~\eqref{eqn:particle-cluster-exact}, 
the difference being that in~\eqref{eqn:particle-cluster-exact} 
the target ${\bf x}_i$ interacts with the source particles ${\bf y}_j$,
while in~\eqref{eqn:particle-cluster-approx-rearranged}
it interacts with the Chebyshev points ${\bf s}_{\bf k}$.
In either case, 
the interactions are independent and can be computed simultaneously;
this is essential to the efficient GPU implementation described in Section~\ref{section:gpu-implementation}.

\subsection{Computing the Modified Charges}
\label{section:interpolation-singularities}
As noted above,
the barycentric form
of the Lagrange polynomial $L_k(x)$ in~\eqref{eqn:1d-barycentric}
has removable singularities at the interpolation points $x = s_{k^\prime}$,
and
these must be treated correctly in computing the 
modified charges $\widehat{q}_{\bf k}$ in~\eqref{eqn:modified-weights}.
In this context,
a removable singularity occurs when a coordinate of a source particle coincides 
with a coordinate of a Chebyshev point,
$y_{j\ell}=s_{k_\ell^\prime}$ for some index $\ell=1, 2, 3$.
As discussed below, when generating the source clusters 
we use the minimal bounding box surrounding the particles, 
thereby guaranteeing that some source particle coordinates 
coincide with some interpolation point coordinates.
The resulting removable singularities are 
handled following the procedure in~\cite{Berrut:2004aa,Wang:2019aa};
when computing the modified charges for a cluster, 
if a source particle coordinate $y_{j\ell}$
coincides with a Chebyshev point coordinate $s_{k_\ell^\prime}$ 
to within a given tolerance,
which we take to be the 
smallest positive IEEE normal double precision floating point number, 
then the condition 
$L_{k_\ell}(y_{j\ell}) = \delta_{k_\ell k_\ell^\prime}$ 
is explicitly enforced in evaluating~\eqref{eqn:modified-weights}.

\subsection{Treecode Description}
\label{section:treecode-algorithm}

{\bf Source Clusters and Target Batches.} 
The treecode begins by constructing a hierarchical tree of source clusters.
The root cluster is the minimal bounding box containing all source particles,
and
the root is recursively divided into child clusters,
where the recursion terminates when a cluster contains $N_L$ or fewer particles.
The cluster division occurs at the midpoint of the three dimensions of the bounding box.
Following the source tree construction,
the treecode constructs a set of localized target batches containing 
$N_B$ or fewer target particles per batch.
The same partitioning routines applied to the source particles 
are applied to partition the target particles into batches.
In the examples shown below,
the targets and sources represent the same set of particles
and we set $N_B=N_L$, 
hence the batches are equivalent to the leaves of the source tree.
In general, the targets and sources may refer to different sets of particles, 
and the batches can be any geometrically localized set of targets.
The effect of target batching on the GPU implementation efficiency is discussed below.

\begin{figure}[htb]
\centering 
\includegraphics[width=0.5\textwidth]{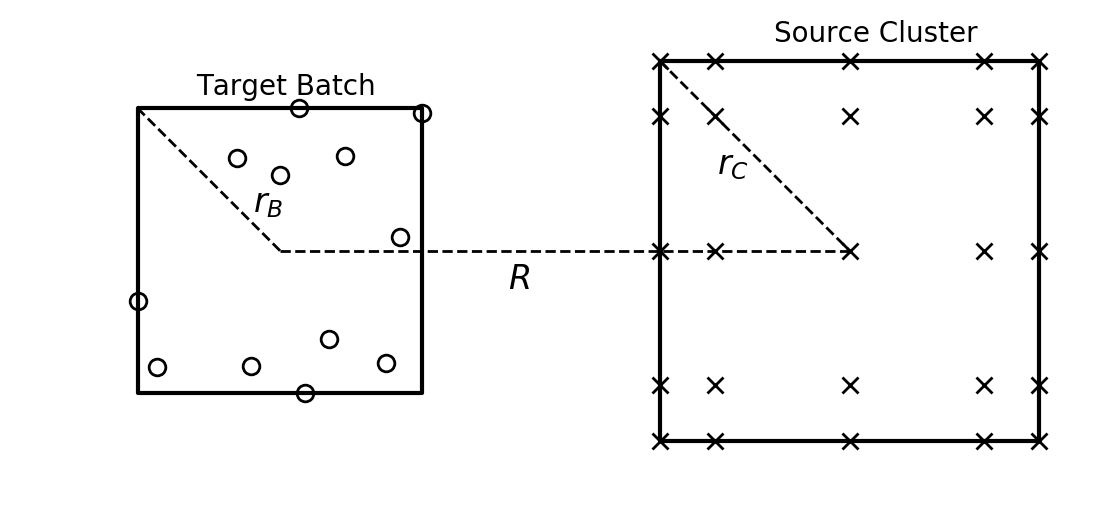}
\caption{2D schematic of target batch/source cluster interaction,
target batch of radius $r_B$ with randomly located target particles ($\circ$),
source cluster of radius $r_C$ with Chebyshev points ($\times$),
batch-cluster distance $R$.}
\label{fig:batch-cluster-interaction}
\end{figure}

{\bf Multipole Acceptance Criterion.} 
The particle-particle interactions~\eqref{eqn:particle-particle} 
are reorganized into target batch/source cluster interactions, 
where the approximation~\eqref{eqn:particle-cluster-approx-rearranged} 
is used if the following multipole acceptance criterion (MAC) ia satisfied, 
\begin{equation} \label{eqn:MAC}
    \frac{r_B + r_C}{R} < \theta, \quad (n+1)^3 < N_C,
\end{equation}
where $r_B$ is the radius of the target batch,
$r_C$ is the radius of the source cluster,
$R$ is the distance between the batch and cluster centers,
$\theta$ is the user-defined MAC parameter,
$n$ is the interpolation degree,
and $N_C$ is the number of source particles in the cluster.
The first condition $(r_B + r_C)/R < \theta$ is diagrammed in Fig.~\ref{fig:batch-cluster-interaction} and ensures the accuracy of the approximation.
The second condition $(n+1)^3 < N_C$ introduces a cluster size check to ensure the efficiency of the approximation.
In particular, 
since the approximate interaction~\eqref{eqn:particle-cluster-approx-rearranged} 
has the same direct sum form as the exact interaction~\eqref{eqn:particle-cluster-exact}, 
if the cluster contains fewer source particles than interpolation points,
$N_C < (n+1)^3$,
it is both faster and more accurate to compute the exact interaction.

{\bf BLTC Algorithm.} 
The BLTC algorithm is given below;
the structure is similar to the original treecode algorithm~\cite{Barnes:1986aa},
however using barycentric Lagrange interpolation
and
target batching as explained above.
The input consists of the particle data and treecode parameters,
and
the output consists of the potential values.
Line 5 constructs the hierarchical tree of source clusters 
and 
the set of localized target batches.
In lines 6-7, the modified charges are computed for each cluster using~\eqref{eqn:modified-weights}.
In lines 8-9, 
each target batch interacts with the root cluster via the recursive function \textsc{ComputePotential}, 
with three options in this function.
If the MAC is satisfied,
then the batch-cluster approximation is computed with \eqref{eqn:particle-cluster-approx-rearranged}.
If the MAC fails because $(r_B+r_C)/R \geq \theta$,
then there are two possibilities: 
if the cluster is a leaf,
then the batch interacts directly with the cluster by~\eqref{eqn:particle-cluster-exact},
otherwise \textsc{ComputePotential} is called recursively for each child of the cluster.
If the MAC fails because $(n+1)^3 \geq N_C$,
then the interaction is computed directly by~\eqref{eqn:particle-cluster-exact}.
The BLTC algorithm requires $O(N\log N)$ operations 
compared to the $O(N^2)$ operations for
direct summation~\eqref{eqn:particle-particle}.
~\\
\hrule
\begin{algorithmic}[1]
\label{alg:treecode}
\Procedure{BLTC}{}
\State \textbf{input}: particle data ${\bf x}_i, {\bf y}_i, q_i$, $i=1,\ldots,N$
\State {\bf input}: treecode parameters $\theta$, $n$, $N_L$, $N_B$
\State {\bf output}: potentials $\varphi_i, i=1,\ldots,N$
\State build tree of clusters $\{C\}$ and set of batches $\{B\}$
\For {each source cluster}
    \State compute modified charges $\widehat{q}_{\bf k}$
    in~\eqref{eqn:modified-weights}
\EndFor
\For{each target batch}
    \State \textsc{ComputePotential}($B$, root\_cluster)
\EndFor
\EndProcedure

\Function{ComputePotential}{batch, cluster}
\If{MAC is satisfied}
    \State compute approximation by~\eqref{eqn:particle-cluster-approx-rearranged}
\ElsIf{$(r_B+r_C)/R \ge \theta$}
    \If{cluster is a leaf}
        \State compute interaction by direct sum in~\eqref{eqn:particle-cluster-exact}
    \Else 
        \For{each child of cluster}
            \State \textsc{ComputePotential}(batch, child)
        \EndFor
    \EndIf
\ElsIf{$(n+1)^3 \ge N_C$}
    \State compute interaction by direct sum in~\eqref{eqn:particle-cluster-exact}
\EndIf
\EndFunction
\hrule
\end{algorithmic}


\section{Implementation Details}
\label{section:implementation-details}

This section gives the details of the 
GPU implementation of the barycentric Lagrange treecode.
The MPI based distributed memory framework is described in Section~\ref{section:mpi-implementation} 
and
the OpenACC based GPU porting is described in Section~\ref{section:gpu-implementation}. 
The resulting implementation runs in parallel on multiple GPU nodes.

\subsection{MPI Implemetation}
\label{section:mpi-implementation}
This section describes the MPI implementation of the BLTC, 
in which one MPI rank is associated with each GPU.
We describe the recursive coordinate bisection (RCB)
used in this work to generate the domain decomposition 
of a particular set of particles,
the distributed memory framework of BLTC based on 
locally essential trees (LET)~\cite{Warren:1992aa},
and 
our MPI implementation of LETs using remote memory access (RMA).  

\begin{figure}
    \centering
    \subfloat[4 partitions]
    {\includegraphics[width=0.2\textwidth]{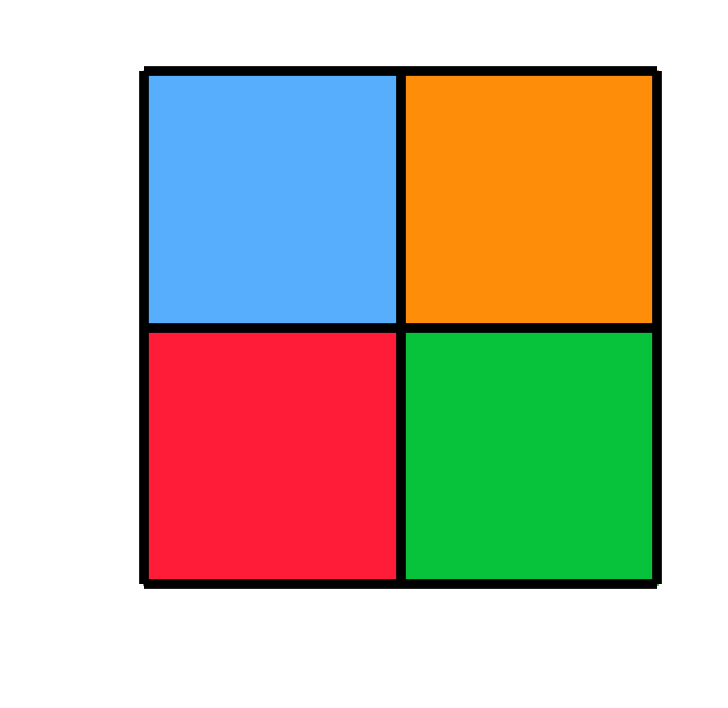}}
    \hfil
    \subfloat[6 partitions]
    {\includegraphics[width=0.2\textwidth]{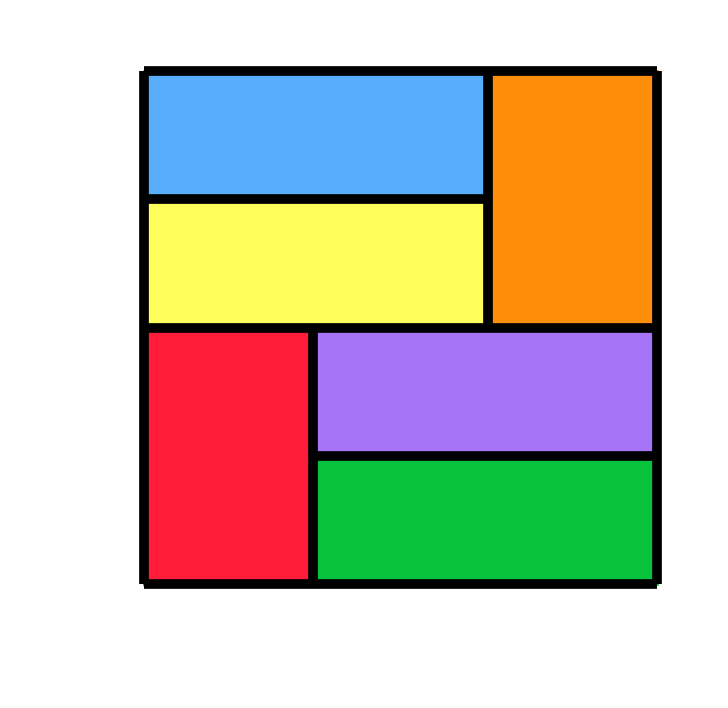}}
    \caption{Recursive coordinate bisection (RCB) of the unit square for 
    (a) 4 partitions, and (b) 6 partitions; 
    coordinate bisection occurs first in the $y$-coordinate 
    and then the $x$-coordinate, repeating until the total number of partitions is reached.
    Each RCB begins with a bisection of the $y$-coordinate at $y=0.5$, 
    assigning half the ranks to the top region and half to the bottom;
    later bisections depend on how many processes were assigned to each partition after the first bisection.
    The area owned by each process in (a) is 1/4, and in (b) is 1/6. }
    \label{fig:RCB}
\end{figure}

{\bf Recursive Coordinate Bisection.}
RCB recursively partitions the domain with a hyperplane that 
(1) is perpendicular to one of the coordinate axes,
and 
(2) balances the number of particles with the number of ranks for each side of the partition.
To construct the test cases below, 
the Zoltan library~\cite{zoltan_v3} is used to perform the RCB.
Figure~\ref{fig:RCB} shows a 2D RCB decomposition of a unit square into four and six partitions, where the area assigned to each partition is balanced.
When building a source tree on the partition, the aspect ratio is taken into account, that is, the ratio of the longest to shortest dimension. 
Typically a cluster is divided into eight children; however, a cluster may be divided into only two or four children if dividing into more would result in aspect ratios greater than $\sqrt{2}$.
For example, the root clusters for each partition in Fig.~\ref{fig:RCB}b would be bisected into two rather than four children.

{\bf Locally Essential Trees.} 
Following the recursive coordinate bisection of the domain, 
each MPI rank owns the particles in one of the partitions, 
and 
it constructs the local source tree for each partition.
The union of local source trees for all partitions is the global source tree, 
but it is never explicitly constructed.
The key observations relevant to LETs are 
(1) each target particle interacts with only a portion of the global source tree, 
and 
(2) nearby target particles interact with similar sub-trees.  
A rank's LET is the union of the interaction sub-trees for all of its target particles.
This LET accounts for all remote data that must be acquired by the rank.
By construction, each LET contains $O(\log N)$ clusters, 
with neighboring ranks exchanging many clusters 
and 
well-separated ranks exchanging few clusters.
Hence, although constructing the LETs requires an all-to-all communication, 
the amount of data acquired by each rank grows only logarithmically 
with the problem size. 

{\bf Remote Memory Access.}
We use MPI passive target synchronization remote memory access (RMA) to perform the construction and communication of the LETs.
First introduced in the MPI-2 standard, 
RMA provides a one-sided communication model within MPI. 
In MPI one-sided operations, 
an origin process can \textit{put} data onto a target process 
or \textit{get} data from a target process 
through specially declared memory \textit{windows}, 
with no active involvement from the target process. 
In active target synchronization, or active RMA, 
the target process sets bounds on when its windows can be accessed; 
in passive target synchronization, or passive RMA, 
the target process puts no limitation on access to its windows, 
and instead the origin process \textit{locks} the target window to perform operations on it. Passive RMA in particular is similar in spirit to the partitioned global address space (PGAS) model used in languages like UPC. In our implementation, we use passive RMA to communicate data between processes.

We believe that one-sided RMA used here is an attractive option for LET construction. 
In the passive RMA approach, we can asynchronously launch all communications with no input from the target rank, and the origin rank need only know the layout of the data to be received from the target. Thus, each rank can construct its LET completely asynchronously from other ranks.

{\bf LET Construction.} 
The LETs are constructed in two steps, which we describe for a two-rank example. 
The tree array (containing cluster midpoints and radii for all tree nodes), 
source particles, 
and cluster charges
are contained within RMA windows on each rank which can be accessed by the other rank.
In the first step, 
rank 1 gets the tree array from rank 2 and creates its interaction lists, 
and vice versa.  
The interaction lists for rank 1
consist of all clusters on rank 2 that a target particle on rank 1 interacts with directly or via the approximation.  
Then, in the second step, rank 1 uses the newly constructed interaction lists to get the necessary source particles and cluster charges from rank 2, filling rank 1's LET.
Simultaneously, rank 2 gets the data it needs from rank 1 to build its LET.
At the conclusion of the second step, 
each rank contains the data needed to perform its calculation, 
and each rank proceeds to compute the potential at its target particles 
as described in the section below on OpenACC implementation.



\subsection{OpenACC Implementation}
\label{section:gpu-implementation}

This section describes the GPU implementation details of the BLTC, 
first discussing target batching, 
followed by
several OpenACC porting details involving host and device memory management, 
GPU compute kernels, and asynchronous streams.
The implementation is extended to multiple GPUs 
on a single node or multiple nodes
in a straightforward manner 
using the MPI implementation described above with one MPI rank per GPU.

{\bf Target Batching.}
Two important factors that affect GPU performance are 
{\it occupancy} and {\it thread divergence}.
Occupancy refers to the fraction of individual hardware compute units that are active at a given time; high occupancy is desired. 
Thread divergence refers to the situation where different software threads follow different logic paths; thread divergence should be avoided.
We achieve high occupancy by batching the target particles and structuring the GPU compute kernels to compute the interactions between all targets in a batch and a source cluster (represented by source particles or interpolation points).  
For large enough batch and leaf cluster sizes 
($N_B,N_L \approx 2000$ for the GPUs used in this work), 
this compute kernel structure achieves high GPU occupancy.
However,
batching would cause thread divergence 
if the targets in the batch interact with different sets of source clusters.
We prevent this by applying the MAC~\eqref{eqn:MAC} to the batch as a whole,
as opposed to applying a MAC to each target particle individually.
While applying the MAC uniformly is sub-optimal for individual targets,
it is nearly optimal because the batch consists of localized target particles; 
moreover the increased GPU performance that comes from avoiding thread divergence
more than compensates.

{\bf Host and Device Data Management.}
The host (CPU) and device (GPU) do not share the same memory,
and
since data movement between them is expensive, 
good memory management is essential for GPU performance. 
In this work,
for a given MPI rank,
all data movement between host and device is managed with OpenACC data regions.
Data transfer to and from the device occurs twice during the calculation.
First, the source particles owned by the rank 
are copied from the CPU onto the GPU,
where the modified charges are computed for each cluster in the rank's sub-tree,
and
then the modified charges are copied back to the CPU's RMA windows 
where other ranks can access them during LET construction.
Second, following LET construction on the host, 
the rank's target particles and LET are copied onto the GPU 
where the potentials are computed,
and
once this is completed for all targets, 
the potentials are copied back to the CPU.
The algorithm below shows the steps for both MPI related communication and host-device communications, labelled HtD for host-to-device and DtH for device-to-host.
\\
\hrule
\begin{algorithmic}[1]
\label{alg:parallel-treecode}
\Procedure{MPI + OpenACC BLTC}{}
\State build tree of clusters $\{C\}$ and set of batches $\{B\}$ from local particles
\State HtD: copy source data
\For {each source cluster}
    \State compute modified charges $\widehat{q}_{\bf k}$
    in~\eqref{eqn:modified-weights} on GPU
\EndFor
\State DtH: copy modified charges
\State create MPI RMA windows to local data
\For{each remote rank}
\State MPI: get tree arrays from remote rank
\EndFor
\State construct interaction lists from tree arrays
\For{each remote rank}
\State MPI: get required particle and cluster data from remote rank and fill into LET 
\EndFor
\State HtD: copy LET
\For{each target batch}
    \State \textsc{ComputePotential}($B$, LET\_root) on GPU
\EndFor
\State DtH: copy final potential
\EndProcedure
\hrule
\end{algorithmic}
~\\

{\bf Overview of Compute Kernels.}
The GPU implementation uses four compute kernels, 
two for preprocessing and two for potential evaluation. 
The preprocessing kernels compute the modified charges~\eqref{eqn:modified-weights}
for each source cluster.
The potential evaluation kernels compute the interaction 
between a target batch and a source cluster, 
either by direct summation~\eqref{eqn:particle-cluster-exact} or 
the cluster approximation~\eqref{eqn:particle-cluster-approx-rearranged}.
The kernels are generated with OpenACC directives, compiled with the PGI C compiler.
For example, we enclose the compute regions with \texttt{\#pragma acc kernels} and identify the parallelizable loops with \texttt{\#pragma acc loop independent}.

{\bf Preprocessing Kernels.}
We describe the two preprocessing kernels 
used to compute the modified charges~\eqref{eqn:modified-weights}.
The first preprocessing kernel computes the intermediate quantities, 
\begin{equation}
    \label{eqn:preproc-kernel-1} 
    \widetilde{q}_j =  \frac{q_j} { \displaystyle
    \sum\limits_{k_1=0}^n \frac{w_{k_1}}{y_{j1}-s_{k_1}} \sum\limits_{k_2=0}^n \frac{w_{k_2}}{y_{j2}-s_{k_2}} 
    \sum\limits_{k_3=0}^n \frac{w_{k_3}}{y_{j3}-s_{k_3}}},
\end{equation}
for each source particle ${\bf y}_j$ in the cluster, 
where each source particle is handled by a single block.
Within each block, the threads parallelize over the interpolation degree $n$, 
computing each term of the three denominator sums 
simultaneously, followed by a reduction.    

The second preprocessing kernel computes the 
modified charges~\eqref{eqn:modified-weights} 
for each Chebyshev point ${\bf s}_{\bf k}$ 
from the intermediate quantities $\widetilde{q}_j$,
\begin{equation}  
    \label{eqn:modified-weights-kernel}
    \widehat{q}_{\bf k} = \sum\limits_{\mathbf{y}_j\in C} \frac{w_{k_1}}{y_{j1}-s_{k_1}}
    \frac{w_{k_2}}{y_{j2}-s_{k_2}}
    \frac{w_{k_3}}{y_{j3}-s_{k_3}}
    \widetilde{q}_j.
\end{equation}
Each Chebyshev point is handled by single thread block. 
Within each block, the threads parallelize over the source particles in the cluster,
computing each term of the sum simultaneously,
followed by a reduction to compute $\widehat{q}_{\bf k}$.  
If the number of source particles exceeds the number of threads per block in the kernel launch,
each thread will be responsible for multiple source particles.  
This also holds for the kernels below where the threads are parallelized over a cluster's source particles or interpolation points.
For interpolation degree $n$ and a cluster containing $N_C$ source particles, 
the first preprocessing kernel performs $O((n+1)N_C)$ operations and the second performs $O((n+1)^3N_C)$ operations for each cluster.

\begin{figure*}
\centering
\subfloat[Direct Interaction Matrix]{
\includegraphics[width=0.3\textwidth]{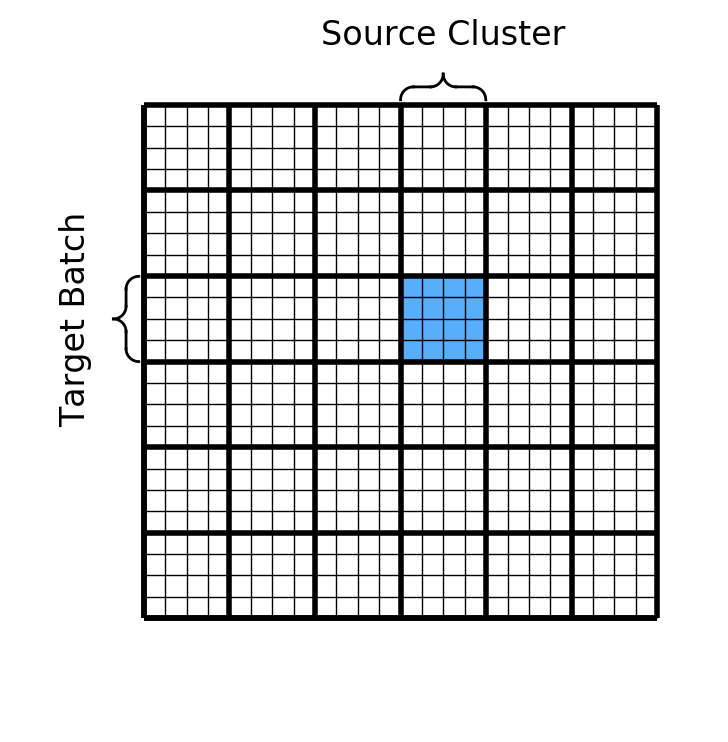}}
\hspace{-10pt}\raisebox{4.5\height}{\LARGE$\xrightarrow{\text{\small Kernel Launch}}$}
\subfloat[Single batch-cluster interaction]{
\includegraphics[width=0.3\textwidth]{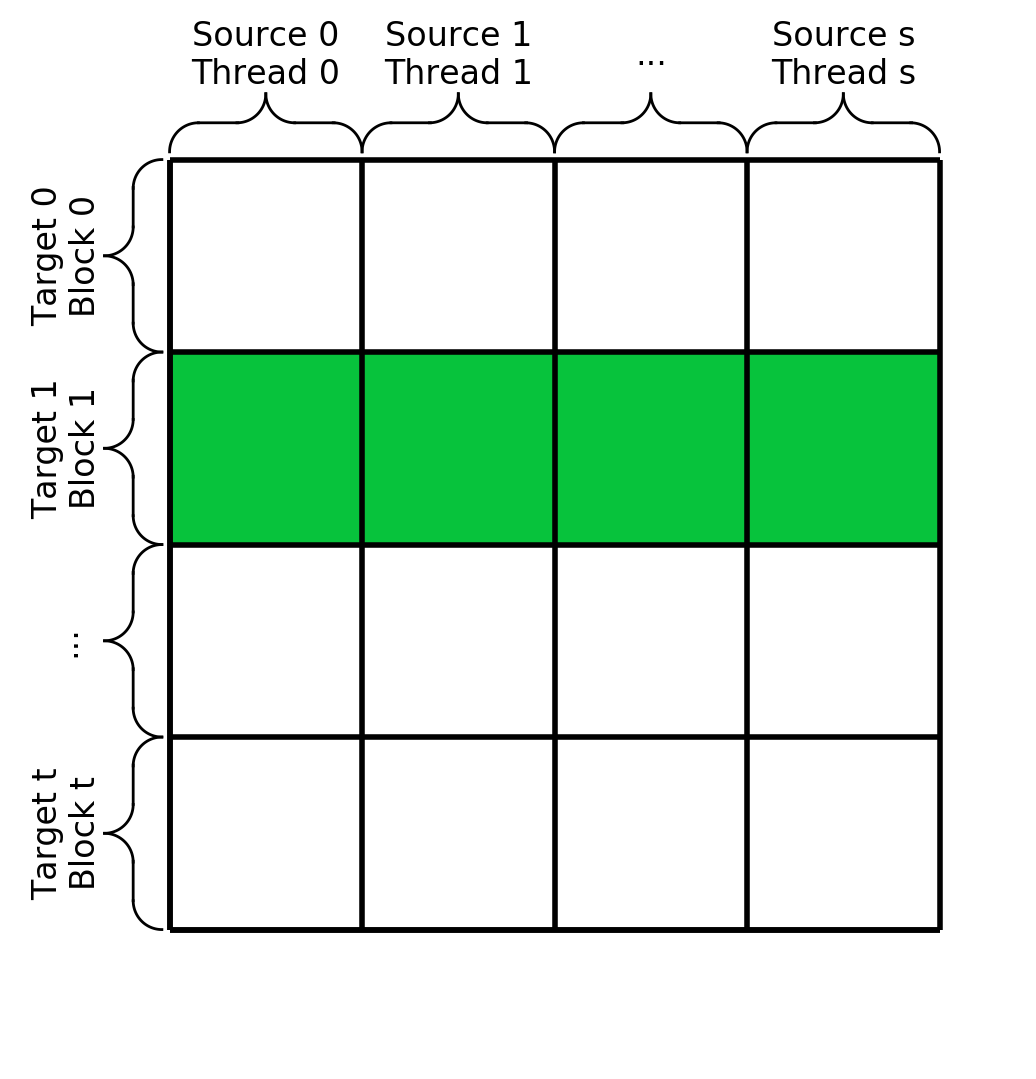}}
\caption{
Structure of batch-cluster direct sum kernel,
a) blue highlighted square represents the interaction between a 
target batch and a source cluster, 
which is computed by a single GPU kernel launch,
(b) green highlighted row represents the interaction between a
target particle and a source cluster,
which is computed by a single thread block.}
\label{fig:batch-cluster-diagram}
\end{figure*}

{\bf Batch-Cluster Direct Sum Kernel.}
This kernel computes the potentials for a batch of target particles 
due to a cluster of source particles by 
direct summation~\eqref{eqn:particle-cluster-exact}.
The kernel is launched when the MAC fails for a leaf cluster.
Figure~\ref{fig:batch-cluster-diagram}a depicts one such kernel launch;
each row represents one of the target particles,
which are organized into batches (bold horizontal partitions),
while each column represents one of the source particles, which are organized into clusters (bold vertical partitions).
Highlighted in blue is the work done by one launch of the batch-cluster direct sum kernel.
Figure~\ref{fig:batch-cluster-diagram}b shows how the work is arranged on the GPU.
The batch-cluster interaction consists of an outer loop over the target particles in the batch 
and an inner loop over the source particles in the cluster.
Since the potentials at different targets are independent, 
the outer loop is naturally parallelizable and we assign each target to a thread block.
One such thread block is highlighted in green in Fig.~\ref{fig:batch-cluster-diagram}b.
The inner loop is parallelized over the threads, with each particle-particle interaction computed by a single thread,
e.g., the $j$th thread in the $i$th block computes $G(\mathbf{x}_i,\mathbf{y}_j)q_j$.
Finally, a reduction over the threads computes the aggregate potential due to all sources in the cluster. 
For a batch containing $N_B$ target particles and a cluster containing $N_C$ source particles, this kernel performs $O(N_CN_B)$ operations.

{\bf Batch-Cluster Approximation Kernel.}
This kernel computes the potentials for a batch of target particles 
due to a cluster of Chebyshev points by the approximation~\eqref{eqn:particle-cluster-approx-rearranged}.  
The kernel is launched when the MAC passes.  
Importantly, 
the approximation~\eqref{eqn:particle-cluster-approx-rearranged}
has the same direct sum structure as the 
exact interaction~\eqref{eqn:particle-cluster-exact}, 
where the inner loop over source particles is replaced by an inner loop over Chebyshev points; 
hence
Fig.~\ref{fig:batch-cluster-diagram} applies to this kernel as well.
For a batch containing $N_B$ target particles, 
this kernel performs $O((n+1)^3N_B)$ operations.
  
The direct sum form of the 
particle-cluster approximation~\eqref{eqn:particle-cluster-approx-rearranged},
with its independent kernel evaluations,
is the distinguishing feature of the 
barycentric Lagrange treecode that permits an efficient GPU implementation.
This form allows each target particle to interact 
simultaneously with the Chebyshev points in a source cluster,
enabling an inner level of parallelization not possible in 
some other approaches such the 
Taylor treecode~\cite{Li:2009aa},
which rely on fundamentally serial recurrence relations to compute 
particle-cluster approximations.


{\bf Asynchronous Streams.}
During the treecode potential evaluation,
the CPU is responsible for looping through the interaction lists 
and launching the GPU kernels.
However when a kernel is launched,  
if the CPU waits to regain control until after the calculation completes,
there are several sources of inefficiency: 
(1) the CPU is idle while the GPU is working,  
(2) there is an initialization cost associated with each kernel launch,
and 
(3) a single kernel might not saturate the GPU with work.
To mitigate these issues we employ asynchronous streams 
(\texttt{\#pragma acc kernels async(streamID)}), 
which allow the CPU to queue the kernel on the GPU 
and 
immediately regain control without waiting for the calculation to complete.
The CPU will then queue the next item from the interaction list 
on a different stream, before the first has completed.
As the code loops through the interaction list,
it cycles \texttt{streamID} through the number of available streams, 
which for the GPUs used in this work is four.

With this approach the initialization on one stream is overlapped 
with the computation on other streams, reducing the GPU idle time.
Furthermore, the GPU may decide to work on multiple streams at the same time if it has available resources, further improving  efficiency.
To handle memory access conflicts and race conditions, 
we use an atomic update (\texttt{\#pragma acc atomic}) 
when updating the potential for a given target particle.
Asynchronous streams significantly improve the performance of our GPU implementation. 
For example, in the 1~million particle test case described in Section~\ref{section:results}, 
asynchronous streams reduce the computation time in a typical case 
by about 25\%.

{\bf Code Availability.}
The code is publicly available on GitHub at github.com/Treecodes/BaryTree,
as both a stand alone executable and a library, with examples.


\section{Numerical Results}
\label{section:results}

We demonstrate the GPU-accelerated barycentric Lagrange treecode 
on a series of test cases 
ranging from 1~million to 1~billion particles.  
In each case the particles are randomly uniformly distributed in the $[-1,1]^3$ cube, 
with charges randomly uniformly distributed on $[-1,1]$.
For these tests, the targets and sources are the same set of particles,
although the code is not restricted to this case.
We present results for the Coulomb and Yukawa potentials ($\kappa=0.5$), 
but the code is fully capable of treating more complex kernels.
Investigation of irregular particle distributions arising from various physical systems is left for future work.
All reported times are the wall clock run time in seconds and 
include the setup phase, precompute phase, and compute phase.
The setup phase includes the data movements and 
communication required for each rank to begin its local calculation;
this consists of organizing the local source particles into an octree 
and target particles into batches, 
construction and communication of the LET, 
and 
creation of the interaction lists for each target batch.
The precompute phase computes the modified charges for each locally owned source cluster, and
the compute phase computes the potential at each target particle.
The calculations are done in double precision arithmetic
and the reported errors are the relative 2-norm error,
\begin{equation}
    E = \left( \sum_{i=1}^N (\varphi_i^{ds} - \varphi_i^{tc})^2 \bigg/ \sum_{i=1}^N (\varphi_i^{ds})^2\right)^{1/2}, 
\end{equation}
where $\varphi_i^{ds}$ are the potentials computed by 
direct summation~\eqref{eqn:particle-particle} 
and $\varphi_i^{tc}$ are computed by the treecode.
For large systems with 8~million or more particles,
the error was sampled at a random subset of target particles.


{\bf Single GPU vs. Single 6-Core CPU.}
We begin by comparing the GPU implementation to a 
portable CPU implementation for a 1~million particle test case.
The computations were performed on the Flux HPC Cluster at the University of Michigan.
The CPU calculations are run on a 6-core 2.67 GHz Intel Xeon X5650 processor 
using OpenMP to perform a straightforward shared-memory parallelization with loop parallelization. The code was compiled with the PGI C compiler v19.1, with the \texttt{-O3} flag.
Each target batch is assigned to one OpenMP thread, 
which then loops over the batch's interaction list.
The GPU calculations are run on a single NVIDIA Titan V.

\begin{figure*}[htb]
\centering
\subfloat[Coulomb potential]
{\includegraphics[width=0.35\textwidth]{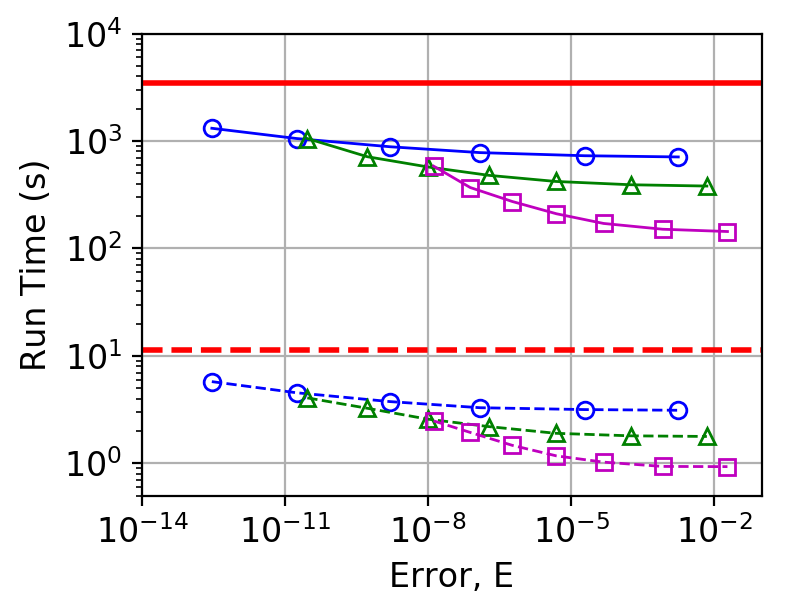}}
\hfil
\subfloat[Yukawa potential]
{\includegraphics[width=0.35\textwidth]{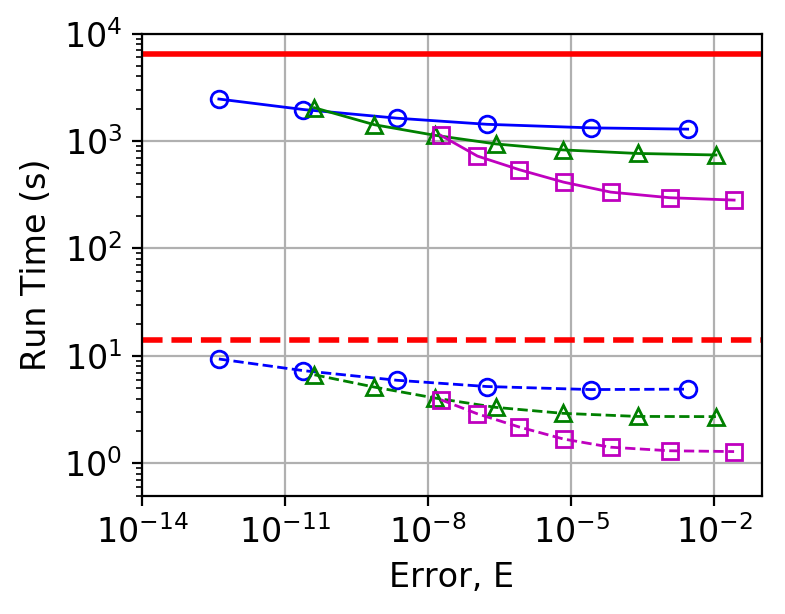}}
\hfil
\hspace{-1cm}
\subfloat  
{\includegraphics[width=0.16\textwidth]{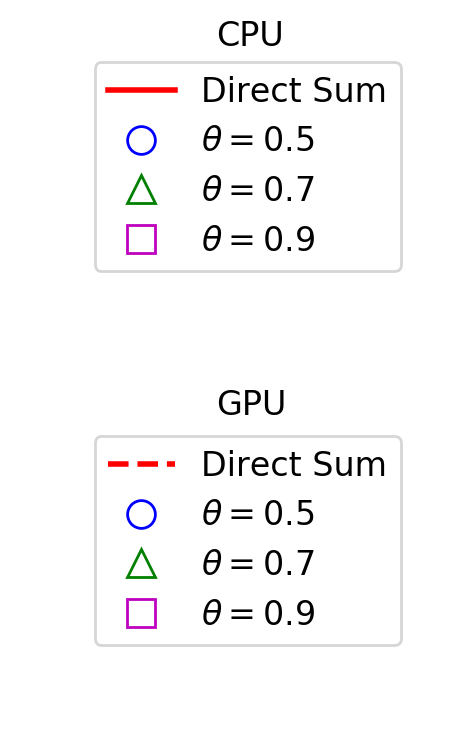}}
\caption[ ]
{Comparison of single GPU and 6-core CPU,
1~million random particles in a cube,
run time versus error, 
(a) Coulomb potential, 
(b) Yukawa potential,
CPU (solid lines/curves), GPU (dashed lines/curves),
direct sum (red horizontal lines),
BLTC results are shown with curves of constant MAC $\theta = 0.5, 0.7, 0.9$,
and
degree $n = 1:2:13$ or until machine precision is reached.}
\label{fig:cpu-gpu-1M}
\end{figure*}

Figure~\ref{fig:cpu-gpu-1M} shows the computation time versus error 
for the (a) Coulomb potential and (b) Yukawa potential.
The batch size and leaf cluster size are $N_B=N_L=2000$.
Each curve represents constant MAC $\theta = 0.5, 0.7, 0.9$, 
where the interpolation degree $n$ is swept from 1 to 13,
or until machine precision is reached,
with solid lines corresponding to CPU results and dashed lines to GPU results.
The red horizontal lines are the run times for direct summation;
note that on the GPU,
the direct sum is computed by one launch of
the batch-cluster direct sum kernel for a batch consisting of all target particles 
and 
a cluster consisting of all source particles.

We draw several conclusions from Fig.~\ref{fig:cpu-gpu-1M};
(1) on both the CPU and GPU, 
the BLTC is faster than direct summation over the entire range of errors
up to machine precision, 
(2) the BLTC runs at least 100$\times$ faster on the GPU than the CPU,
(3) the results for the Coulomb and Yukawa potentials are qualitatively similar,
the main difference being that the run times for the Yukawa potential are slightly higher (approximately $1.8\times$ on the CPU and $1.5\times$ on the GPU),
(4) while the GPU direct sum is faster than the CPU treecode for this problem size, 
this will not be the case for large enough problems due to the 
$O(N^2)$ scaling of direct summation.

\begin{figure}[htb]
\centering
\includegraphics[width=0.49\textwidth]{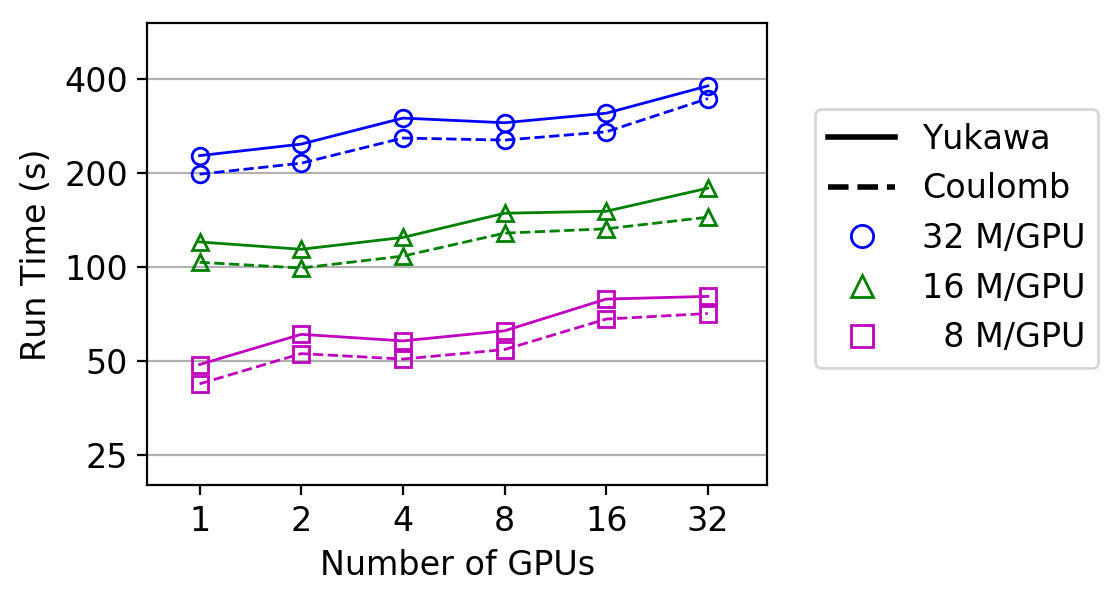}
\caption{Weak scaling of GPU-accelerated BLTC on Comet,
parameters MAC $\theta = 0.8$, degree $n=8$, batch/leaf size $N_L=N_B=4000$
yielding 5-6 digit accuracy,
run times for Coulomb potential (dashed lines) 
and Yukawa potential (solid lines) versus number of GPUs,
number of particles per GPU = 8, 16, 32~million, 
number of GPUs = 1:32,
largest system is 1.024 billion particles
(345~s for Coulomb, 380~s for Yukawa).}
\label{fig:weak-scaling}
\end{figure}

{\bf Weak Scaling.}
We demonstrate weak scaling of the GPU implementation 
by holding the number of particles per GPU fixed
and 
increasing the number of GPUs from 1 to 32.
The calculations were performed on Comet using NVIDIA P100 GPUs; 
our XSEDE startup allocation allows up to eight nodes containing four P100s per node. The code was compiled with the PGI C compiler v18.10 and Open MPI v4.0.2, with the \texttt{-O3} flag.
The treecode parameters are MAC $\theta=0.8$, degree $n=8$, 
and
batch/leaf size $N_L=N_B=4000$, yielding 5-6 digit accuracy.
Figure~\ref{fig:weak-scaling} shows the run time for the 
Coulomb potential (dashed lines) and Yukawa potential (solid lines),
with the number of particles per GPU set to 8, 16, and 32~million 
(squares, triangles, circles).
For the largest test case with $N=1.024$ billion particles, 
the Coulomb potential run time was 345~s with error 7.6e-6,
and
the Yukawa potential run time was 380~s with error 1.5e-5.
Note that the run times increase only modestly as the problem size grows;
this is consistent with the $O(N\log N)$ scaling of the BLTC.
In some instances the time plateaus as the number of ranks increases, 
e.g. between 4 and 8 GPUs for 32 million particles/GPU;
this is attributed to variations in the domain decomposition and the resulting leaf and batch sizes. 


{\bf Strong Scaling.}
We demonstrate the strong scaling of the GPU-accelerated BLTC
on Comet using up to 32 NVIDIA P100 GPUs. 
The test systems consist of 16 million and 64 million particles 
interacting via the Coulomb and Yukawa potentials.
We use the same BLTC parameters as above,
MAC $\theta=0.8$, degree $n=8$, batch/leaf size $N_L=N_B=4000$,
yielding error 4.0e-6 and 5.9e-6 for the Coulomb potential
and
3.0e-6 and 7.2e-6 for the Yukawa potential, respectively for the two system sizes.

\begin{figure*}[ht!]
\centering
\subfloat[Coulomb potential, efficiency]
{\includegraphics[width=0.35\textwidth]{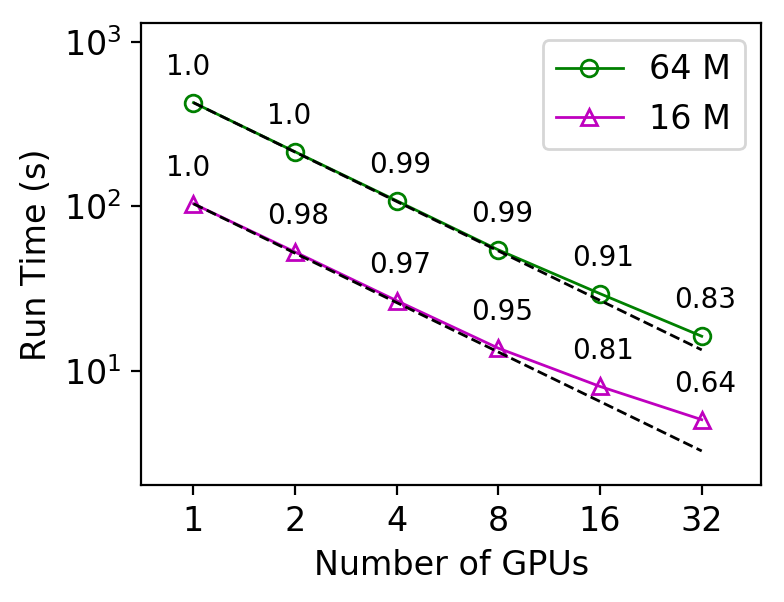}}
\hfil
\subfloat[Yukawa potential, efficiency]
{\includegraphics[width=0.35\textwidth]{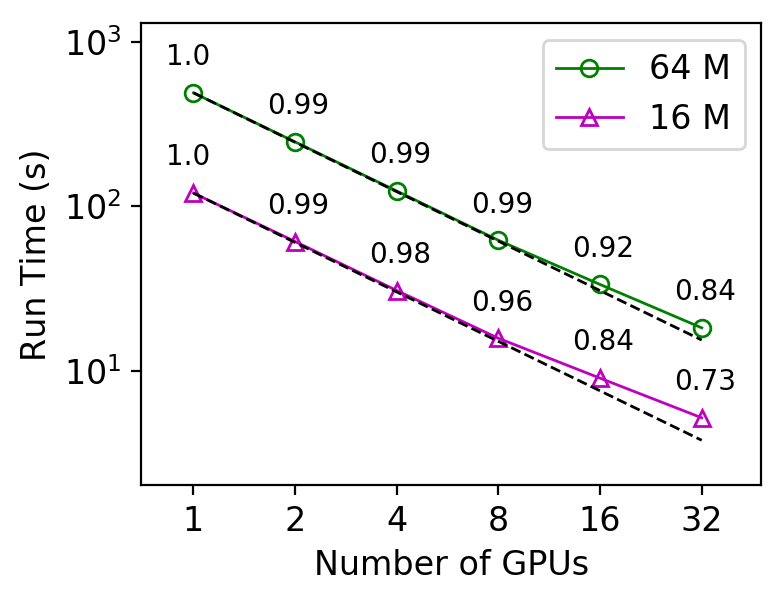}}
\hfil
\subfloat[Coulomb potential, time distribution]
{\includegraphics[width=0.35\textwidth]{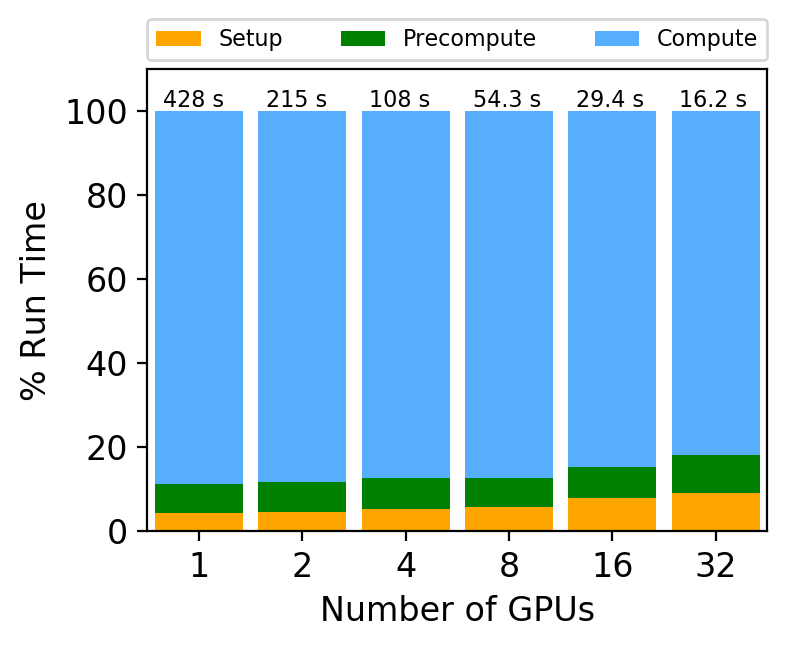}}
\hfil
\subfloat[Yukawa potential, time distribution]
{\includegraphics[width=0.35\textwidth]{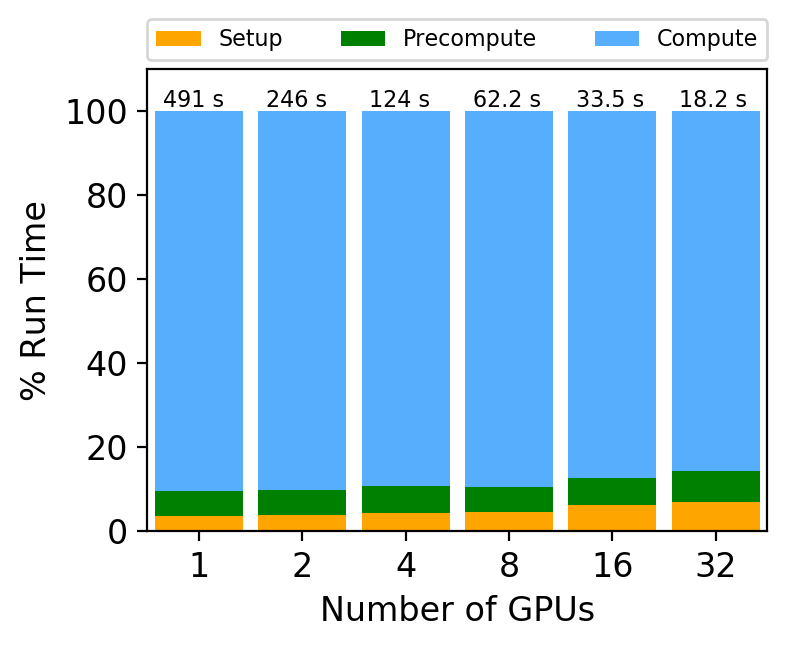}}
\caption{
Strong scaling of GPU-accelerated BLTC
on Comet with up to 8 nodes (32 GPUs), 
BLTC parameters MAC $\theta = 0.8$, degree $n=8$, batch/leaf size $N_B=N_L=4000$,
(a,c) Coulomb potential,
(b,d) Yukawa potential,
(a,b) 16M and 64M particles, run time labeled with efficiency relative to a single GPU,
error 4.0e-6, 5.9e-6 (Coulomb) and 3.0e-6, 7.2e-6 (Yukawa),
(c,d) 64M particles, percent of time spent in setup, precompute, and compute phases 
as the number of GPUs increases, total time above each bar.
}
\label{fig:strong-scaling-coulomb}
\end{figure*}

Figure~\ref{fig:strong-scaling-coulomb}ab
shows the strong scaling efficiency of these calculations 
by plotting the run time versus the number of GPUs.
The efficiency is measured with respect to a single GPU 
and 
is compared to ideal speedup (dashed lines).
As the number of GPUs is increased to 32, 
the 64M particle example maintains higher efficiency (83\%, 84\%) 
than the 16M particle example (64\%, 73\%). 

Figure~\ref{fig:strong-scaling-coulomb}cd 
shows the distribution of time spent in each phase of the calculation 
as the number of GPUs increases from 1 to 32 for the 64M particle example, 
distinguishing between the setup phase (orange),
the precompute phase (green),
and the compute phase (blue).
For a given number of GPUs,
the bar is colored based on the percent of run time spent in each phase, 
with the total time listed above the bar.  
Up to 32 ranks (with 2M particles/rank) the compute phase dominates the total time, 
but as the number of GPUs increases, the work shifts towards the setup 
and 
precompute phases. 
The fraction of time spent in the setup phase grows because the communication costs grow; 
more interactions are with remotely owned data that must be communicated.
The fraction of time spent in the precompute phase grows because the modified charge kernels do not saturate the GPUs with work as the number of particles per rank decreases.
Nonetheless,
with 64 million particles on 32 GPUs,
the BLTC requires 16.2~s for the Coulomb potential
and
18.2~s for the Yukawa potential to achieve 5-6 digit accuracy.


\section{Conclusions}
\label{section:conclusion}
We presented an MPI + OpenACC implementation of the 
barycentric Lagrange treecode (BLTC) for 
fast summation of particle interactions on GPUs.
The BLTC relies on 
barycentric Lagrange interpolation at Chebyshev points of the 2nd kind
to approximate well-separated particle-cluster interactions,
and
it is kernel-independent because it requires only kernel evaluations.
The distributed memory parallelization 
uses recursive coordinate bisection for domain decomposition 
and MPI remote memory access to build locally essential trees on each rank.
The particle interactions are organized into
target batch/source cluster interactions which efficiently map onto the GPU;
target batching provides an outer level of parallelism,
while the direct sum form of the barycentric particle-cluster
approximation~\eqref{eqn:particle-cluster-approx-rearranged}
provides an inner level of parallelism.

The first test was performed on an NVIDIA Titan V GPU.
We demonstrated significant speedup of the GPU-accelerated BLTC 
over a 6-core CPU counterpart for a system consisting of
1~million randomly distributed particles interacting via the 
Coulomb and Yukawa potentials,
for a range of errors up to machine precision as the 
interpolation degree $n$ increases.

Subsequent tests were performed on Comet using NVIDIA P100 GPUs,
where the BLTC parameters were set to achieve 5-6 digit accuracy.
We investigated parallel scaling of the BLTC on up to 8 GPU nodes 
with 4 GPUs per node.
We demonstrated good weak scaling for a fixed number of particles per rank, 
showing only a moderate increase in run time as the number of ranks increased
from 1 to 32,
consistent with the $O(N\log N)$ scaling of the BLTC.
The largest calculation had 32 million particles per rank, 
on 32 ranks, for a total of 1.024 billion particles,
where the run times were 345~s for the Coulomb potential
and
380~s for the Yukawa potential.
We demonstrated good strong scaling for systems with 16 million and 64 million particles;
with 64 million particles on 32~GPUs, 
the BLTC requires 16.2~s for the Coulomb potential
and
18.2~s for the Yukawa potential,
maintaining about 83\% parallel efficiency in both cases.

In future work we will investigate techniques for improving efficiency such as overlapping communication and computation and using mixed-precision arithmetic.
We will extend the GPU implementation to the barycentric Hermite treecode, 
which is similar to the BLTC, but requires low order partial derivatives of the kernel~\cite{Krasny:2019aa}.
We will also explore GPU acceleration of 
barycentric cluster-particle and cluster-cluster 
treecodes~\cite{Appel:1985aa,Deng:2012aa, Boateng:2013aa}.
In addition,
the GPU-accelerated BLTC presented here is being applied to 
Poisson--Boltzmann continuum solvation computations~\cite{Geng:2013aa}, 
and
density functional theory electronic structure calculations~\cite{vaughn-gavini-krasny-2020}. 


\bibliographystyle{IEEEtran}
\bibliography{IEEEabrv,library}

\end{document}